\documentclass[showpacs,preprintnumbers,amsmath,amssymb,floatfix,eqsecnum]{revtex4}
\usepackage{graphicx}% Include figure files
\usepackage{dcolumn}% Align table columns on decimal point
\usepackage{bm}% bold math
\usepackage[]{natbib}
\usepackage{subfigure}
\begin{document}

%\preprint{APS/}
\title{High-frequency behavior of w-mode pulsations of compact stars}
% Force line breaks with \\

\author{Y. J. Zhang\footnote{Present address: Department of Physics,
University of California at San Diego, 9500 Gilman Drive, La
Jolla, CA 92093, USA}}

\author{J. Wu\footnote{Present address: Department of Physics and
Astronomy, University of Pittsburgh, Pennsylvania 15260, USA. }}
\author{P. T. Leung\footnote{Email:
ptleung@phy.cuhk.edu.hk}}
\affiliation{%
Physics Department and Institute of Theoretical Physics, The
Chinese University of Hong Kong, Shatin, Hong Kong SAR, China.
}%

\date{\today}% It is always \today, today,
\def\tomega{ \tilde{\omega} }
\def\tOmega{ \tilde{\omega} }
\def\tr{ \tilde{r} }
\def\tx{ \tilde{r}_* }
\def\tV{ \tilde{V} }
\def\tpsi{ \tilde{\psi} }
\def\trho{ \tilde{\rho} }
\def\tP{ \tilde{P} }
\def\tR{ \tilde{R} }
\def\tX{ \tilde{R}_* }
\def\tm{ \tilde{m} }
\def\tphi{ \tilde{\nu} }
\def\tlam{ \tilde{\lambda}}
\def\tepsilon { \tilde{\epsilon}}
\def\tU{ \tilde{U} }
\def\tphi{ \tilde{\nu} }
\def\tlam{ \tilde{\lambda}}
\def\tepsilon { \tilde{\epsilon}}
\def\cc{{\cal C}}
\def\N{{\cal N}}
\def\R{{\cal R}}
\def\a{ {\rm a}}
\def\c{ {\rm c}}
\def\e{ {\rm e}}
\def\p{ {\rm p}}
\def\f{ {\rm f}}
\def\d{ {\rm d}}
\def\i{ {\rm i}}
\def\r{ {\rm r}}
\def\A{ {\rm A}}
\def\P{ {\rm P}}
\begin{abstract}
We study the asymptotic behavior of the quasi-normal modes (QNMs)
 of w-mode pulsations of compact stars in the high-frequency regime. We observe that
 both the axial and polar w-mode QNMs attain similar asymptotic behaviors in spite of the
 fact that they are described by two totally different differential equation systems.
 We obtain robust asymptotic formulae relating w-mode QNMs of
 different polarities and different angular momenta. To explore the physical reason underlying such similarity, we first derive
 a high-frequency approximation for the polar w-mode oscillations
 to unify the descriptions for both cases. Then, we
 develop WKB-type analyses for them and quantitatively explain
 the observed  asymptotic behaviors  for polytropic stars and quark stars.
 We also point out that such asymptotic behaviors  for realistic stars
 are strongly dependent on the equation of state near the stellar
 surface.
\end{abstract}

% PACS, the Physics and Astronomy Classification Scheme.
%\keywords{gravitational waves --- stars: neutron ---  stars:
%oscillations (including pulsations) --- equation of state ---
%relativity

\maketitle
\section{Introduction}
%\chapter{Asymptotic Behavior of Spacetime Modes}
Gravitational waves (GWs) are undoubtedly a major speculation of
general relativity that has not yet been observed directly.
Decades of efforts of physicists from different disciplines have
been pooled together to design and build GW detectors of various
designs (see, e.g., \cite{Hughes_03} and references therein). To
date several Earth-based GW interferometric detectors including
GEO600, LIGO, TAMA300 and VIRGO have been operating. In spite of
the fact that these  detectors are not yet sensitive enough to
detect GWs directly, interesting upper limits have been placed on
either the GW strains or event rate for several potential
astrophysical sources (see, e.g.,
\cite{LIGO_S4_BHringdown,LIGO_S5_LowMass,PhysRevD.79.122001,PhysRevD.82.102001}
for the results obtained from the latest science runs of LIGO).
Optimistically speaking, GWs could be observable and analyzed
directly within one or two decades when more advanced detectors
are available.

A promising mechanism for GW emission in astrophysics is achieved
through pulsating  blackholes (BHs) and neutron stars (NSs). When
these compact stellar objects are perturbed away from their
equilibrium states in various astronomical processes, such as
BH-BH, BH-NS and NS-NS binary mergers,   and asymmetric stellar
core collapse (see, e.g.,
\cite{Belczynski:2001uc,Hughes_03,Lindblom:1998wf,Fryer:2001zw}),
huge amounts of energy can be released in the form of GW, which
might be detectable with our current (or future) GW observatories.
While quantitative studies on such fierce processes demand
sophisticated techniques in numerical relativity (see, e.g.,
\cite{yamamoto:064054,dimmelmeier:064056,baiotti:084033,duez:104015,
etienne:044024,kiuchi:064037,abdikamalov2009} for some recent
studies in this field), it is also popular to study the dynamics
of the end products of these processes, usually in the form of
pulsating BHs or NSs, with linearized theory, which may be able to
throw light on analyses of numerical data gathered at the final
stage of numerical simulations (see, e.g., \cite{baiotti:024002}
for a comparison between linear and nonlinear treatments of
pulsating NSs).

As in other linear oscillating systems, relevant physical
quantities in pulsating BHs or NSs and the associated GW signals
are assumed to have $\exp(-i \omega t)$ time dependence, where
$\omega$ is the eigenfrequency of the system under consideration.
However, due to energy carried away by GWs, these quantities
decay exponentially in time, leading to complex-valued frequency
$\omega \equiv \omega_\r+i\omega_\i$ with $\omega_\i \leq 0$.
Correspondingly, these oscillation modes are termed as
quasi-normal modes (QNMs)
\cite{Press_1971,Leaver_1986,rmp,Kokkotas_rev,Berti:2009kk}. QNMs
of compact stellar objects are interesting because they can, to
certain extent, reflect the physical characteristics (e.g., mass,
radius, moment of inertia, equation of state (EOS) and
composition) of the GW emitter. In particular, a host of efforts
have been paid to infer the structure of a NS, which is often
masked by uncertainties in EOS of nuclear matter or quark matter,
from its QNM spectra (see, e.g.,
\citep{Andersson1996,Andersson1998,Ferrari,Kokkotas_2001,
Ferrari_prd,Tsui05:3,Tsui:prd,lau-2009}). On the other hand, it
has been conjectured that the asymptotic behavior of  highly
damped QNMs (i.e., modes with large $|\omega_\i|$) of BHs is
related to a calculation of the Bekenstein entropy in loop quantum
gravity and to the quantum of area, thus spurring theorists to
consider these modes with lively interest (see, e.g.,
\cite{Berti:2009kk,maggiore:141301} and references therein).

Motivated by the physical significance of the asymptotic behavior
of BH QNMs, we study in the present paper  the high-frequency
asymptotic behavior of the (spacetime) w-mode QNMs of NSs.
Although the axial and polar w-mode QNMs are described by two
totally different differential equation systems,  we find that
they acquire similar asymptotic behaviors and there exist robust
asymptotic formulae relating these two kinds of QNMs.
 To theoretically understand these rather surprising results, we first derive
 a high-frequency approximation (HFA) for the polar w-mode oscillations
 to unify the descriptions for polar and axial w-mode oscillations of NSs. Subsequently, we
 develop WKB-type analyses for these QNMs and quantitatively explain
 the observed  asymptotic behaviors  for polytropic stars and quark stars.
 We also show how these asymptotic behaviors of w-mode QNMs in realistic stars
 could depend on the EOS near the stellar surface.

The organization of this paper is summarized as follows. In
Sect.~II we present the observed asymptotic behavior of axial and
polar w-mode QNMs of NSs, which is totally different from that of
BHs. We also show how these two kinds of w-mode QNMs of NSs are
related in the high-frequency regime. In Sect.~III we introduce
the HFA for polar w-mode oscillations and hence show that axial
and polar w-mode QNMs are governed by two similar second-order
differential systems, setting the stage for a unified treatment
for the asymptotic behaviors for the two kinds of w-mode. In
Sect.~IV we establish a WKB-type analysis for relevant QNMs and
derive the asymptotic behavior for w-modes of polytropic stars and
quark stars described by the MIT bag model (see, e.g.,
\cite{ComStar}). Through our analysis developed in Sect.~IV, we
can explain our findings summarized in Sect.~II and show how such
asymptotic behavior is related to the EOS of a star. In Sect.~V we
propose a symmetric and unified way to describe high-frequency
axial and polar w-mode oscillations on equal footing. In Sect.~VI
we consider the case of realistic stars and discuss several key
factors affecting our analysis. We then conclude our paper with a
brief discussion and some remarks on the difference between the
asymptotic behaviors of QNMs of BHs and NSs in Sect.~VII.

Unless otherwise stated explicitly, geometric units where $G=c=1$
are adopted throughout the whole paper, and kilometers are used as
the unit of lengths. Besides, we will only consider non-rotating
NSs (with ideal fluids) and BHs in the present paper. As usual,
the geometry of a static gravitational source with spherical
symmetry (NS or BH) is described by the line element:
\begin{eqnarray}
ds^2 = g_{\alpha\beta}dx^\alpha dx^\beta = -e^{\nu}
dt^2+e^{\lambda}dr^2+r^2(d\theta^2+\sin^2\theta d\phi^2).
\end{eqnarray}
For NSs the metric coefficients $e^{\nu(r)}$ and $e^{\lambda(r)}$
can be obtained from the equilibrium configuration  governed by
the Tolman-Oppenheimer-Volkov (TOV) equations
\cite{Tolman:1939jz,Oppenheimer:1939ne}.

\section{Asymptotic Behavior of w-mode QNMs}
The oscillation modes of both NSs and BHs can be classified into
polar and axial classes according to the property of the metric
perturbations under inversion. For typical stellar models, the
polar class includes (fundamental) f-mode, (gravity) g-mode,
(pressure) p-mode and (spacetime) w-mode (see, e.g.,
\cite{Kokkotas_rev} for a review). Unlike the f-, g- and p-modes
which have their Newtonian origins \cite{cox}, the w-modes mainly
represent
 oscillations in spacetime and are specific to theory of general relativity.
 In contrast to the rich spectra of polar oscillations, axial
oscillations of NSs do not introduce non-trivial motion of
 matter and all are classified as the w-mode.

Axial QNMs  of NSs are governed by a second-order differential
equation, namely the Regge-Wheeler equation for NSs
\cite{RWeq,Chandrasekhar1}:
\begin{equation}
\left[\frac{d^2}{d
r_*^2}+\omega^2-V_{\a}(r_*)\right]\psi_{\a}(r_*)=0,
\label{axial-waveeq}
\end{equation}
where the tortoise radial coordinate $r_*$ is related to the
circumferential  radius $r$ through the relationship:
\begin{equation}
r_*(r) = \int_0^r e^{[\lambda(r')-\nu(r')]/{2}}dr',
\label{tortoise-r-in}
\end{equation}
and the NS Regge-Wheeler potential is given by:
\begin{eqnarray}
V_{\a}(r_*) = \left\{ \begin{array}{lr}
\displaystyle{\frac{e^{\nu}}{r^3}}\left[l(l+1)r+4\pi
r^3(\rho-p)-6m\right], & \hspace{0.5cm} r<R; \\ \\
              \displaystyle{\frac{(r-2M)}{r^4}}\left[ l(l+1)r-6 M\right], & \hspace{0.5cm} r \ge
              R.
\end{array} \right. \label{V_rw}
\end{eqnarray}
Here $M$, $R$, $\rho(r)$  and $p(r)$ are the total mass, the
radius, the density and the pressure of the star,   $l$ is the
angular momentum index, and
\begin{equation}\label{mass}
 m(r) \equiv 4 \pi \int_0^r \rho (r') r'^2 dr'\,.
\end{equation}
Note that for $r \ge R$ the NS Regge-Wheeler potential is
identical to that of the BH case.

In contrast to the simplicity of axial oscillations, polar
oscillations of NSs are more complicated as
 oscillations in spacetime and fluid are in general coupled together.
 A set of coupled fifth-order equations governing non-radial
 pulsations of compact stars
were first derived in 1967 by Thorne and Campolattaro
\cite{Thorne}. Based on \cite{Thorne}, Lindblom and Detweiler
reduced the system into a fourth-order one
\cite{Lindblom_1983,Lindblom-1985}. Subsequently, there have been
several alternative formulations for polar oscillations of compact
stars (see, e.g.,
\cite{Chandrasekhar1,Chandrasekhar2,Chandrasekhar3,Chandrasekhar4,Price,AAKS98,Lindblom-1997}),
 and most of them (if not all) are fourth-order systems.

\begin{figure}[ht]
\centering \subfigure[]{\label{fig1a}
\includegraphics[height=8.5cm]{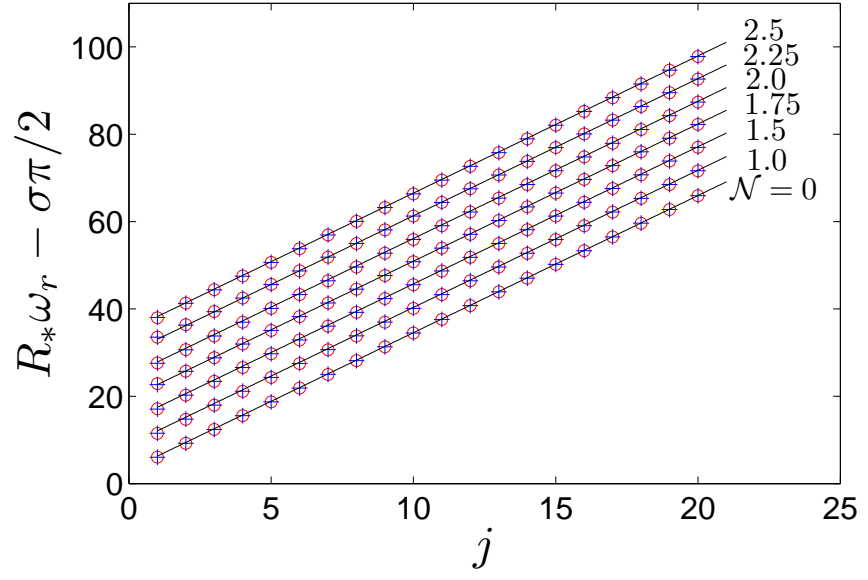}}
\subfigure[]{\label{fig1b}\hspace{1.1cm}
\includegraphics[height=8.6cm]{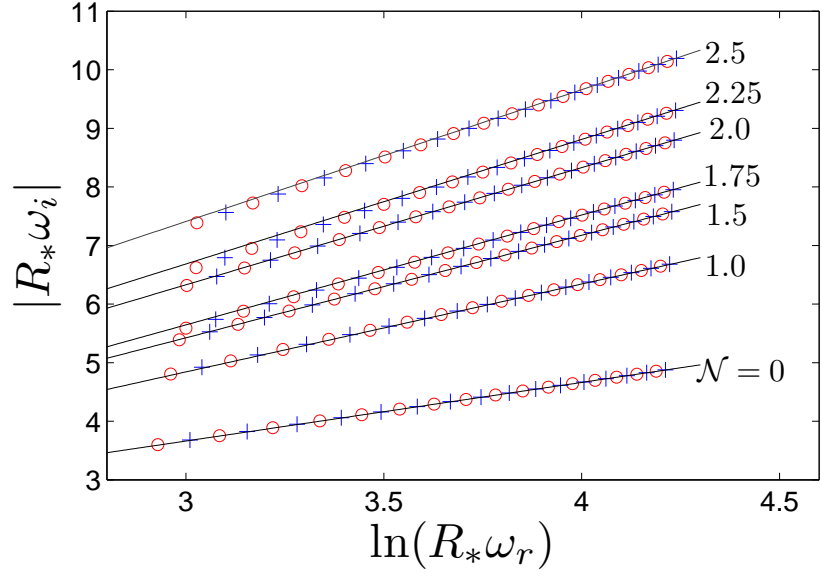}}
\caption{The axial (circles) and the polar (crosses) w-mode QNMs
of the polytropic stars considered in Table~\ref{QSNSPB} are
shown. (a) $R_* \omega_{\rm r} -\sigma \pi/2 $ is plotted against
the mode number $j$, $j=1,2,3,...,20$. For clarity the data of
each polytropic star are upward-shifted by 5, 10, 15, 20, 25, 30
for $\N=1,1.5,1.75,2.0,2.25,2.5$, respectively. (b) $|R_*
\omega_{\rm i}|$ is plotted against $\ln(R_* \omega_{\rm r})$ for
$5 \leq j \leq 20$. The solid straight lines in (a) and (b) are
the corresponding theoretical lines predicted by (\ref{asymp-w-r})
and (\ref{asymp-w-i}), respectively.}
 \label{fig1}
 \end{figure}

 As introduced above, axial and polar oscillations are governed by
 two completely different formalisms. Therefore, it seems
plausible  that axial and polar w-mode QNMs would acquire
different asymptotic behaviors. We have carried out numerical
calculations to locate these QNMs for various stars, including
polytropic stars (with EOS $p \propto \rho^\gamma$), quark stars
(with EOS $p=(\rho-4B)/3$) and NSs constructed with several
popular EOSs of nuclear matter, to examine relevant high-frequency
behavior. The results obtained are in fact a bit surprising and
encouraging as well. As shown in Fig.~\ref{fig1}, where axial and
polar w-mode QNMs of quark stars  and polytropic stars are shown
and analyzed, for a specific star these two kinds of w-mode QNMs
lie on a common smooth curve star in the complex $\omega$-plane,
which can be approximated by the following equations:
\begin{eqnarray}
\omega_{\r} R_* & = & \pi j- \frac{(\widetilde{\theta}_b-\sigma \pi)}{2}, \label{re} \\
\omega_{\i} R_* & = &
\frac{1}{2}\ln\left[r_{b}(R_*)^{\N+2}\right]-(\frac{\N}{2}+1)\ln(\omega_{\r}R_*),\label{im}
\end{eqnarray}
where $j=1,2,3,\ldots$ is the mode number labeling QNMs in order
of increasing frequency,  $\widetilde{\theta}_b$ and $r_{b}>0$ are
real parameters dependent on the star, $\sigma=0 (1)$ for axial
(polar) w-mode, $\N \equiv 1/(\gamma -1)$ is the so-called
polytropic index ($\N=0$ for quark stars hereafter), and $R_*$ is
the tortoise radius of the star.

The situation for realistic stars is less uniform  and it is
difficult to extract a generic asymptotic behavior for $\omega_\i$
(see Fig.~\ref{fig2}). However, $\omega_\r$ still obeys an
approximate equation similar to (\ref{re}), namely
\begin{equation}\label{RS}
\omega_\r \widetilde{R}_* = \pi j -
\frac{(\widetilde{\theta}_b-\sigma \pi)}{2}.
\end{equation}
Here the parameter $\widetilde{R}_*$ is usually a bit less than
$R_*$ and could have a mild dependence on $\omega$. Despite the
lack of an explicit formula for $\omega_\i$, there is a universal
relationship between the complex QNM frequencies of axial and
polar w-modes:
\begin{equation}
\omega_j^{(\P)}=(\omega_j^{(\A)}+\omega_{j+1}^{(\A)})/2
\label{uni}
\end{equation}
and vice versa, where the superscripts P and A denote polar and
axial modes, respectively. This equation holds asymptotically for
large $j$ and works nicely irrespective of EOS of the star
(including various nuclear matter EOSs, polytropic EOS and quark
matter EOS). The main objective of the present paper is to
establish a proper theory to explain these observed asymptotic
behaviors, which will be the focus of the following sections. In
fact, we will show how the values of $\widetilde{\theta}_b$ and
$r_{b}$ for polytropic stars and quark stars can be obtained
analytically.

\section{Wave equation for polar w-mode}
The  similarity between the asymptotic behaviors of the two kinds
of w-mode is a bit surprising since they are in general governed
by two entirely different sets of equations. The findings
discussed in the last section strongly hint that axial and polar
w-modes could actually  be placed on an equal footing at least in
the large-frequency limit. In this section we shall develop a HFA
for polar w-modes under the inverse-Cowling approximation (ICA)
\cite{in_Cowling,wu07} to show that they can be described by a
single second-order wave equation analogous to
(\ref{axial-waveeq}). To this end, we first introduce an
alternative formalism of non-radial polar oscillations of compact
stars suggested by Allen, Andersson, Kokkotas and Schutz
\cite{AAKS98}, referred to as AAKS formalism in the following
discussion.

\subsection{AAKS Formalism}
In AAKS formalism  polar oscillations of compact stars are
described by two spacetime metric variables $F$, $S$ and one
fluid perturbation variable $H$:
\begin{eqnarray}
rF & = & h_{\theta\theta},\\
rS & = & h_{tt}-\frac{e^{\nu}}{r^2}h_{\theta\theta},\\
H & = & \frac{\delta p }{\rho+p} \label{H-define},
\end{eqnarray}
where $h_{\alpha\beta}$ ($\alpha,\beta=t,r,\theta,\phi$) denotes
the first-order perturbation in the metric,  and $\delta p$ is the
Eulerian variation in pressure \cite{AAKS98}. These three physical
quantities satisfy three second-order evolution equations plus one
second-order time-independent (or equivalently
frequency-independent) Hamiltonian constraint equation and one can
use any three of these four equations to solve for $F$, $S$ and
$H$. In this paper we shall use the two wave equations:
\begin{eqnarray}
&&\omega^2 S+\frac{d^2 S}{d r_*^2}+\frac{2e^{\nu}}{r^3}\left[ 2\pi
r^3(\rho+3p)+m-(n+1)r\right]S
=-\frac{4e^{2\nu}}{r^5}\left[ \frac{(m+4\pi p r^3)^2}{r-2m}+4\pi \rho r^3-3m\right]F\label{Allen-St}\,,\\
&&\omega^2 F+\frac{d^2F}{d r_*^2}+\frac{2e^{\nu}}{r^3}\left[ 2\pi
r^3(3\rho+p)+m-(n+1)r\right]F=-2\left[ 4\pi
r^2(p+\rho)-e^{-\lambda}\right]S+8\pi(\rho+p)re^{\nu}\left(
1-\frac{1}{C_s^2}\right)H\label{Allen-Ft}\,,\nonumber\\
\end{eqnarray}
 plus the Hamiltonian constraint:
\begin{eqnarray}
0& = & \frac{d^2F}{d r_*^2}-\frac{e^{(\nu+\lambda)/2}}{r^2}(m+4\pi r^3p)\frac{d F}{d r_*}+\frac{e^{\nu}}{r^3}[12\pi r^3\rho-m-2(n+1)r]F \nonumber \\
& & -re^{(\nu+\lambda)/2}\frac{d S}{d r_*}+\left[ 8\pi
r^2(p+\rho)-(n+3)+\frac{4m}{r}\right]S+\frac{8\pi
r}{C_s^2}e^{\nu}(\rho+p)H  \label{constraint}
\end{eqnarray}
to develop a second-order wave equation for polar w-mode.  In
these equations $n = (l-1)(l+2)/2$, and $C_s$ is  the sound speed
in the stellar fluid.

\subsection{Inverse-Cowling approximation}
It is commonly accepted that polar w-mode oscillations are
oscillations in spacetime and fluid elements are to certain extent
spectators. Hence, it is reasonable to expect that polar w-mode
QNMs could be reproduced under the so-called ICA where certain
fluid perturbations (e.g. fluid displacements or pressure
variation) are neglected \cite{in_Cowling,wu07}. In fact, it has
been shown that ICA assuming zero Lagrangian pressure variation
can produce quantitatively correct result \cite{wu07}. Here we
propose another version of ICA, namely zero Eulerian pressure
variation (i.e., $\delta p=0$), to derive a valid wave equation
for polar w-mode QNMs.

From (\ref{Allen-Ft}) and (\ref{constraint}), we have:
\begin{equation}
8\pi re^{\nu}\delta p =
\omega^2F+\frac{e^{(\nu+\lambda)/2}}{r^2}(m+4\pi
pr^3)\frac{dF}{dr_*}+\frac{e^{\nu}}{r^3}(3m+4\pi
pr^3)F+re^{-(\nu+\lambda)/2}\frac{dS}{dr_*}+(n+1)S.
\label{delta_p}
\end{equation}
Under our ICA scheme, $\delta p=0$, and consequently
Eq.~(\ref{delta_p}) can be rewritten in the following way:
\begin{equation}
F=-\frac{e^{(\nu+\lambda)/2}(m+4\pi pr^3)}{\omega^2 r^2}\frac{dF}{
dr_*}-\frac{e^{\nu}(3m+4\pi pr^3)}{\omega^2 r^3}F-\frac{r
e^{-(\nu+\lambda)/2}}{\omega^2}\frac{dS}{dr_*}-\frac{n+1}{\omega^2}S.
\label{delta_p0}
\end{equation}
This equation, together with (\ref{Allen-St}), can lead to a
single wave equation in $S$.

\subsection{High-frequency approximation}
We note that the frequencies of w-mode are at least a few times
higher than that of $f$-mode, which is proportional to
$(M/R^3)^{1/2}\sim\overline{\rho}^{1/2}$, with $\overline{\rho}$
being the average density of the star. As a result, terms of the
order $\overline{\rho}^{1/2}/\omega$ are small for w-mode QNMs and
can be neglected for high-order modes. Based on this observation
and (\ref{delta_p0}), we have developed a systematic power
expansion in $1/\omega$ for $F$. Hence, the contribution of $F$ to
(\ref{Allen-St}) can be included perturbatively, leading to an
accurate scheme to solve for $S$. The details of such perturbation
scheme are provided in Appendix~A for reference. However, since we
are interested in the high-frequency asymptotic behavior of polar
w-mode, we adopt the lowest order HFA where  the contribution of
$F$ to (\ref{Allen-St})  is neglected completely to obtain a
single wave equation for $S$:
\begin{equation}\label{HFA}
\left[\frac{d^2}{d r_*^2}+\omega^2-V_{\rm p}(r_*)\right]S(r_*)=0
\,
\end{equation}
with
\begin{eqnarray}
V_{\rm p}(r_*) = \left\{ \begin{array}{lr}
\displaystyle{\frac{e^{\nu}}{r^{3}}}\left[l(l+1)r-4\pi
r^3(\rho+3p)-2m\right], & \hspace{0.5cm}r<R; \\
\displaystyle{\frac{e^{\nu}}{r^{3}}}\left[l(l+1)r-2M\right], &
\hspace{0.5cm}r \ge R.
\end{array} \right.
\end{eqnarray}
This polar w-mode wave equation is identical to
 the wave equation for the axial mode, (\ref{axial-waveeq}), except for
 the functional form of the potential term. We will see in the following discussion that such similarity
 between the governing equations of these two kinds of w-mode wave
 is in fact the origin of the peculiar asymptotic behavior observed in
 Sect.~II.

To gauge the accuracy of HFA and examine the validity of
(\ref{HFA}), we located the first ten polar w-mode QNMs with
(\ref{HFA}) and compared the numerical results with  the exact
values in Tables \ref{HFA-PS} and \ref{HFA-RS} for a polytropic
star (EOS: $p=100\rho^2$) with a compactness $0.297$, and  a
realistic star (EOS: APR2 \cite{APR}) with a compactness $0.196$,
respectively. The numerical results of $\omega$ are given in units
of $M^{-1}$.  From the numerical results, it is obvious that the
error of HFA decreases with increasing $\omega$. Therefore, the
HFA can indeed provide us  accurate results for high-order modes.

\section{Asymptotic Behavior of w-modes for model
Stars}\label{AB-axial} In the above discussion we have shown that
both axial and polar w-mode QNMs of compact stars are governed by
a Klein-Gordon type wave equation:
\begin{equation}
\left[\frac{d^2}{d r_*^2}+\omega^2-V_{}(r_*)\right]\psi_{}(r_*)=0,
\label{KG-waveeq}
\end{equation}
where $V=V_{\a}\, (V_{\rm p})$ and $\psi_{}=\psi_{\a}\,(S)$ for
axial (polar) case. Despite the fact that the form of
(\ref{KG-waveeq}) is similar to the Regge-Wheeler equation and the
Zerilli equation describing GW propagation around  BHs
\cite{RWeq,Zerilli}, the potential $V$ has non-trivial dependence
on density and pressure. To see how such dependence could affect
the asymptotic behavior of relevant QNMs, in this section we focus
our attention on stars with simple analytic EOS, namely polytropic
stars and quark stars described by the simplest MIT bag model with
zero strange quark mass.

\subsection{Locating QNMs}
One usual way to locate QNMs is as follows. For any specific
$\omega$, we first find a function $f(\omega, r_*)$ satisfying the
regularity condition $f \sim r_*^{l+1}$ as $r_* \rightarrow 0$ and
a function $g(\omega, r_*)$ satisfying the outgoing condition $g
\sim \exp(i \omega r_*)$ as $r_* \rightarrow \infty$. QNMs are
then located when the Wronskian of $f(\omega, r_*)$ and $g(\omega,
r_*)$ vanishes, i.e.,
\begin{eqnarray}
W(f,g;\omega)\equiv f(\omega, r_*)\frac{d}{d r_*}g(\omega,
r_*)-g(\omega, r_*)\frac{d}{d r_*}f(\omega, r_*)=0,
\end{eqnarray}
and the corresponding eigenfunction is proportional to $f(\omega,
r_*)$ (or equivalently $g(\omega, r_*)$).

In the high-frequency domain  the asymptotic expression of
$f(\omega, r_*)$, $g(\omega, r_*)$ and $W(\omega)$ can be derived
under the standard WKB approximation, and hence an approximate
expression of QNM frequencies can be derived as well
\cite{gw1,gw2}. As in other applications of the WKB
approximation, due attention has to be placed on the singular
points of the potential, e.g., points at which the potential (or
the derivatives of the potential) are discontinuous. For
polytropic stars and quark stars, $V(r_*)$ is well behaved except
at the stellar surface. We shall show in the following analysis
how the reflection of GW at the stellar surface could dominate
the asymptotic behavior of w-mode QNMs.

\subsection{WKB analysis}
Adopting the methods discussed in
\cite{Berry-1982,Slavyanov-1996}, we have developed a systematic
WKB analysis for both $f$ and $g$ (see \cite{Zhang_thesis} for
details). In the high-frequency limit, the dominant parts in $f$
and $g$ are given respectively by
\begin{equation}
f_{}(r_*, \omega \rightarrow \infty)=\omega r_*j_l(\omega r_*),
\hspace{1.0cm}  0 \leq r_* \leq R_*,
\end{equation}
and
\begin{eqnarray}\label{gfunction}
g_{}(r_*, \omega \rightarrow \infty) = \left\{ \begin{array}{lr}
\exp\left[i \omega (r_*-R_*)\right]+\R
\exp\left[-i\omega (r_*-R_*)\right], & 1/\omega \ll r_* < R_*; \\
(1+\R)\exp\left[i\omega(r_*-R_*)\right], & R_* \leq r_* < \infty.
\end{array} \right.
\end{eqnarray}
Here $\R$ is the standard reflection coefficient as defined in
\cite{Berry-1982}. In particular, if the potential $V(r_*)$ has a
discontinuity in $d^N V/d r_*^N$ ($N=0,1,2,\ldots$), $\R=a
\omega^{-(N+2)}$, with
\begin{equation}\label{diff}
a=-\left(\frac{i}{2}\right)^{N+2} \lim_{\varepsilon \rightarrow
0^{+}}\left[\frac{d^N V_{}(r_*)}{d
r_*^N}|_{r_*=R_*+\varepsilon}-\frac{d^N V_{}(r_*)}{d
r_*^N}|_{r_*=R_*-\varepsilon}\right].
\end{equation}
Detailed derivation of the above result can be found in
\cite{Berry-1982}.

The asymptotic eigenvalue equation for w-mode QNMs is then
obtained by setting the Wronskian of $g_{}(r_*,\omega\rightarrow
\infty)$ and $f_{}(r_*,\omega\rightarrow \infty)$ to zero:
\begin{eqnarray}\label{eigen}
b \omega^{-(N+2)}=\exp({-i 2 \omega R_*}),
\end{eqnarray}
where $b\equiv r_{b}\exp(i\theta_{b})=(-1)^{l+1}a$ with $r_{b}>0$
and $\theta_{b}$ being two real constants. This equation can be
solved readily, yielding
\begin{equation}
\omega_{j}\approx \frac{1}{2 R_*} \left\{2 \pi
j-\theta_{b}-i\left[(N+2)\ln \frac{2\pi j-\theta_{b}}{2R_*}-\ln
r_{b}\right]\right\}, \label{wj}
\end{equation}
where $j\gg 1$ is a positive integer and the approximation
$|\omega| \simeq  |\omega_\r|$ has been used.

It is obvious that Eq.~(\ref{wj}) indeed leads to the asymptotic
behavior of w-mode summarized in Sect.~II as long as $N$ is equal
to the polytropic index $\N$. In the following discussion we shall
show that this is the case for $\N=0,1,2,\ldots$. More
importantly, Eq.~(\ref{wj}) can be generalized to include cases
with non-integral values of $\N$. Before diving into detailed
calculation of $b$, several useful features of (\ref{wj}) are
worthy of remark. Firstly, $\omega_\r$ of the $j$-th and
$({j+1})$-th eigenfrequencies are spaced by $\pi/R_*$. Secondly,
$\omega_\i$ depends linearly on the logarithm of $\omega_\r$. The
above features both hold for axial and polar high-frequency
w-modes. In addition, since the leading discontinuity in the
potential $V_{\a}$ or $V_{\rm p}$ is usually caused by the $\pm 4
\pi e^{\nu} \rho$ term, the expression of $b$ has a minus sign
difference between the axial and the polar cases. The phases of
$b$ for the axial and polar cases thus differ by $\pi$, leading to
a shift of $\pi/(2 R_*)$ in $\omega_\r$. All these features are
consistent with our numerical results presented in Sect.~II.

\subsection{Application to model stars}
To make our discussion concrete and physical, we apply our
analysis to quark stars and polytropic stars, which are described
by EOSs with analytic forms. It is readily shown that the
potential and the derivatives of the potential for such stars are
continuous except at the stellar surface. In the following we will
show how the values of $N$ and $b$ can be expressed in terms of
the physical parameters of the star.

For a quark star with the MIT Bag Model EOS ($\rho=3p+4B$, with
$B$ being the bag constant), the star surface is defined at the
point where the pressure vanishes. However, the density  there is
equal to $4B$ and does not vanish. Therefore, the potential has a
direct discontinuity at the surface of the star, i.e., the $N=0$
limit of (\ref{diff}), and straightforward calculation shows that
\begin{eqnarray}
b & = &
%\pm(-1)^{l}(-\frac{1}{4})\lim_{\varepsilon\rightarrow 0^{+}}\left[V(R_*+\varepsilon)-V(R_*-\varepsilon)\right] \\
%  & = &
(-1)^{l+\sigma}4B \pi \left(1-\frac{2M}{R}\right).
\label{coef-b-qs}
\end{eqnarray}

We now turn our attentions to polytropic stars with EOS given
explicitly by $p=\kappa \rho^{1+1/\N}$, where $\kappa$ is a
constant. Again, it is easy to see from TOV equations that the
variables $m$, $\rho$, $p$ and $\nu$ of polytropic stars are all
analytic functions inside the star and hence our theory applies.
Near the surface of the star, following directly from the TOV
equations, the density and the pressure behave respectively as
\cite{Lindblom_1983}:
\begin{eqnarray}
p & = & \alpha (R_*-r_*)^{\N+1}, \label{p-poly}\\
\rho & = & \beta (R_*-r_*)^\N,   \label{rho-poly}
\end{eqnarray}
where
\begin{eqnarray}
\beta &= &\left[\frac{1}{\kappa (\N+1)}\frac{M}{R^2}\right]^\N,\\
\alpha& =&\kappa \beta^{1+1/\N}.
\end{eqnarray}
We first consider a simple case where $\N$ is equal to  a positive
integer $N$. Then the lowest order discontinuity appears in the
$N$-th derivative of the potential at the surface and the
coefficient $b$ can be found explicitly:
\begin{equation}
b=(-1)^{l+\sigma} \pi
\Gamma(\N+1)\left(1-\frac{2M}{R}\right){\left[\frac{-iM}{2\kappa
(\N+1)R^2}\right]}^{\N}, \label{coef-b}
\end{equation}
where $\Gamma(z)$ is the  gamma function. It is interesting to
note that Eq.~(\ref{coef-b}) in fact includes (\ref{coef-b-qs}) as
the special case $\N=0$.

It follows directly from the asymptotic analysis developed above
that for $j \gg 1$
\begin{eqnarray}
\omega_{\r} R_* & = & \pi j-\frac{\left(\widetilde{\theta}_b-\sigma \pi\right)}{2}, \label{asymp-w-r}\\
\omega_{\i} R_* & = & -(\frac{\N}{2}+1)\ln(\omega_r
R_*)+\frac{1}{2}\ln\left[r_{b}(R_*)^{\N+2}\right],
\label{asymp-w-i}
\end{eqnarray}
with
\begin{equation}\label{theta_b}
 \widetilde{\theta}_b\equiv {\theta}_b+\sigma \pi=\left(-l-\frac{\N}{2}\right)\pi \,.
\end{equation}
It is obvious that the forms of   (\ref{asymp-w-r}) and
(\ref{asymp-w-i}) agree with the numerical results  summarized in
Sect.~II. Most interestingly, Eqs.~(\ref{coef-b}),
(\ref{asymp-w-r}) and (\ref{asymp-w-i}) can be readily
generalized to non-integral values of $\N$ by method of analytic
continuation. Thus, we have established a generic and unified
asymptotic behavior for polytropic stars and quark stars as well.

\subsection{Numerical results}
 To further examine the validity and accuracy of our
method, we have calculated the exact eigenfrequencies for w-mode
QNMs of several model stars (including one quark star and six
polytropic stars whose physical parameters are tabulated in
Table~\ref{QSNSPB} for reference) and showed the results in
Fig.~\ref{fig1}. It is clearly shown in these figures that the
asymptotic behavior of w-mode QNMs agrees nicely with the
theoretical results derived above. For example, the spacing in
$\omega_\r$ between two successive QNMs of the same polarity is
close to $\pi/R_*$. There is one axial (polar) QNM at the midpoint
of the line joining two neighboring polar (axial) QNMs (i.e.
Eq.~(\ref{uni})). Besides, $\omega_\i$ depends linearly on $\ln
j$. These features are just the direct consequence of
(\ref{asymp-w-r}) and (\ref{asymp-w-i}).

On the other hand, the coefficient $b$ for each case can be
calculated from (\ref{coef-b}) with the physical parameters
tabulated in Table~\ref{QSNSPB}. Substituting the so-obtained
value of $b$ (see Table~\ref{QSNSPB})
 into (\ref{asymp-w-r}) and (\ref{asymp-w-i}), we can
determine both $\omega_{\r}$ and $\omega_{\i}$ for $j \gg 1$
without using any unknown free paramter. To further verify our
theory, we have fitted exact values of $\omega_{\r}$ and
$\omega_{\i}$ with the following functions:
\begin{eqnarray}
\omega_{\r} R_* & = & c_1 j+c_0, \label{fitting-quark-r}\\
\omega_{\i} R_* & = & d_1\ln(\omega_{r} R_*)+d_0,
\label{fitting-quark-i}
\end{eqnarray}
with $c_1$, $c_0$, $d_1$ and $d_0$ being adjustable parameters to
yield the best fit. The theoretical values of these parameters
(see (\ref{coef-b}), (\ref{asymp-w-r}) and (\ref{asymp-w-i})) and
the values obtained from the best fit to the w-modes (from the
$11$-th to the $20$-th modes) are listed in Tables~\ref{QSNS-C}
(values of $c_1$, $c_0$) and \ref{QSNS-D} (values of $d_1$,
$d_0$). It is obvious that our analysis works nicely in all cases.
The numerical values of both $c_1$ and  $d_1$ are close to their
theoretical ones. For the case of $c_0$ and  $d_0$, the errors are
larger, especially for the former. This is expected as small
errors in the slope would result in much larger errors in the
intercept in the process of extrapolation. Therefore, as a
remedial action, we have first set $c_1$ and $d_1$ to their
theoretical values and substituted the eigenfrequency of the
$20$-th w-mode into (\ref{fitting-quark-r}) and
(\ref{fitting-quark-i}) to compute $c_0$ and $d_0$ directly (the
numeric results enclosed in brackets in Tables~\ref{QSNS-C} and
\ref{QSNS-D}). It is clear that this modified method can improve
the accuracy to certain extent.

\section{Unified approach to w-modes}
From the asymptotic analysis on w-modes developed above we see
that the asymptotic behaviors of w-modes are mainly determined by
the most singular part in the potential, which is responsible for
reflection of GWs at high frequencies. Apart from the trivial
centrifugal barrier that has been duly considered by adopting the
solution $f \propto j(\omega r_*)$,  there are three other terms
in the potential which are proportional to $\rho$, $p$ and $m$,
respectively. As usual, the effect of $p$ is much less than that
of $\rho$. Moreover, $m$ is an integral over the $\rho$ term (see
(\ref{mass})) and should be less singular than $\rho$. Hence, it
is reasonable to argue that both the contributions of $p$ and $m$
to the potential are small at high frequencies. Once the terms
proportional to $p$ and $m$ are neglected in the potential,
$V_{\a}$ and $V_{\rm p}$ are almost identical except for a sign
difference in the term proportional to $\rho$. More importantly,
such argument should work irrespective of the EOS of a star. We
therefore propose to use the following unified approach to
w-modes:
\begin{equation}
\left[\frac{d^2}{d r_*^2}+\omega^2-V(r_*)\right]\psi(\omega,
r_*)=0, \label{asymp-equation}
\end{equation}
with
\begin{eqnarray}
V(r_*) = \left\{ \begin{array}{lr}
\displaystyle{\frac{e^{\nu}}{r^{3}}}\left[l(l+1)r+(-1)^\sigma 4\pi
r^3\rho\right], & \hspace{0.5cm}r<R; \\\\
\displaystyle{\frac{1}{r^{3}}}l(l+1)(r-2M), & \hspace{0.5cm}r \ge
R.
\end{array} \right. \label{V-r-star}
\end{eqnarray}
 Thus, the two kinds of
w-modes are cast in a symmetrical description.

We have tested the validity of our proposal by comparing the
numerical results of the w-mode QNM eigenfrequencies obtained from
the above approximate wave equation  with the exact values in
Tables~\ref{Axial-w} (axial case) and \ref{Polar-w} (polar case).
The star under consideration is a realistic NS with EOS A
\citep{modelA}, which has a compactness of $0.27$ and a central
density $\rho_c=2.227\times 10^{-3}$. The w-modes listed in the
tables are the $3$-rd to $12$-th modes. One can see that the
unified approach works accurately for both kinds of w-mode. The
percentage error in $\omega_\r$ and $\omega_\i$ is less than $1\%$
in most cases and decreases with increasing order. Thus, the
validity of the approximate form of the potential in
(\ref{V-r-star}) is justified. As a result, w-modes of both
polarities can be cast into a unified description.

\section{Generalization to Realistic Neutron Stars}
\subsection{Numerical result}
It is not easy to generalize the method established above to
analyze the w-modes of  NSs characterized by realistic EOSs of
nuclear matter for the following reasons. First, such EOSs are
usually given in the form of discrete data points and methods of
interpolation are needed in order to generate the function
$p(\rho)$ (see, e.g., \cite{Arnett1977}). It is likely that
artificial discontinuities in the potential (or its derivatives)
may develop when $p(\rho)$ so generated is used as the input to
solve the TOV equations describing the static configuration of a
compact star.  These artificial discontinuities could then lead to
extra reflections of GW and obscure numerical results of
high-order QNMs and the related theoretical study as well.
Second, the polytropic index of nuclear matter is in general
density-dependent. However, as shown in the previous section, the
polytropic index is a crucial factor in the asymptotic behavior
of w-mode QNM. Except for GWs with very high frequencies,
reflections occurring at interfaces other than the stellar
surface are characterized by different polytropic indices,
leading to more uncertainties in the asymptotic behavior of
realistic NSs.

\begin{figure}[ht]
\centering \subfigure[]{\label{fig2a} \hspace{0.8cm}
\includegraphics[height=8.5cm]{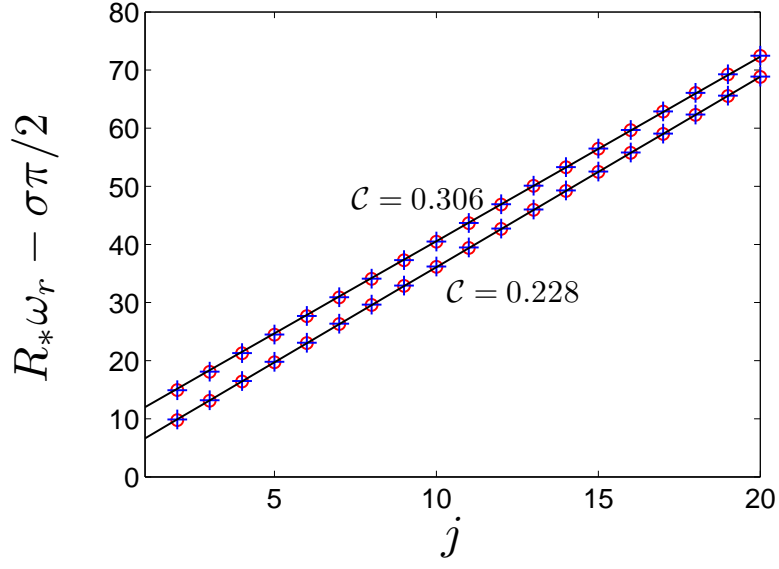}}
\subfigure[]{\label{fig2b}
\includegraphics[height=8.6cm]{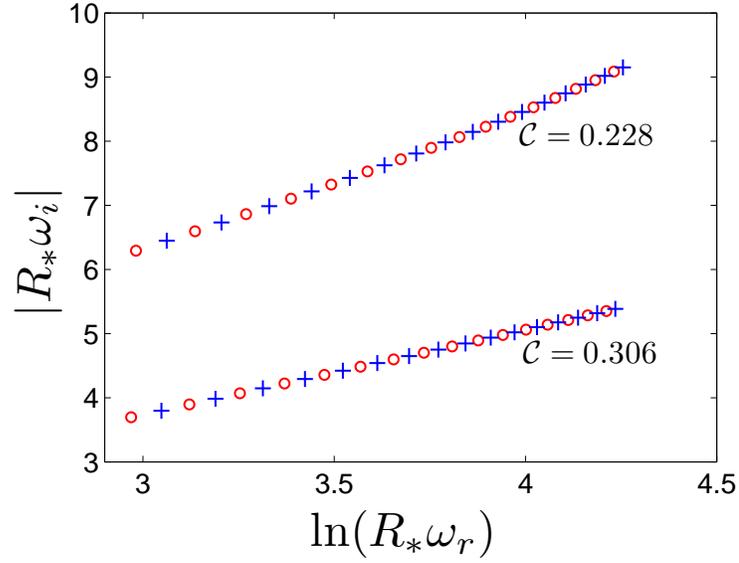}}
\caption{The axial (circles) and the polar (crosses) w-mode QNMs
of two realistic stars obeying the analytic approximation of the
SLy EOS proposed in Ref.~\cite{anaEOS} with compactness ${\cal
C}=0.306,\,0.228$ are shown. (a) $R_* \omega_{\rm r} -\sigma \pi/2
$ is plotted against the mode number $j$, $j=1,2,...,20$. For
clarity for the case ${\cal C}=0.306$ the data points are all
shifted upward by 5. The two straight solid lines shown in the
figure are the best linear fits to the two sets of data,
respectively. (b) $|R_* \omega_{\rm i}|$ is plotted against
$\ln(R_* \omega_{\rm r})$  for $5 \leq j \leq 20$.  }
 \label{fig2}
\end{figure}

Instead of deriving the asymptotic behaviors of w-modes for
realistic compact stars from first principles, we have gathered
some empirical rules for high-frequency w-mode QNMs based on
numerical studies carried out for numerous stellar models,
including models A \cite{modelA}, C \cite{modelC}, three models
(AU, UU and UT) proposed in \cite{AU}, and models APR1 and APR2
proposed in \cite{APR}. In particular, we have constructed NSs
based on the two analytic EOS proposed by Haensel and  Potekhin
\cite{anaEOS}, which  can well approximate FPS EOS \cite{FPS} and
SLy EOS \cite{DH2001} of realistic neutron-star matter. More
importantly, these analytic EOS are free from artificial
discontinuities due to interpolation of discrete data points and
can facilitate theoretical analysis. As a representative example
to illustrate our discussion, we show in Fig.~\ref{fig2} the
w-mode QNMs of two NSs obeying  the analytic SLy EOS  with
compactness ${\cal C}=0.306,\,0.228$, from which one can see the
following features. First, despite the fact that $\omega_\r$ no
longer follows (\ref{asymp-w-r}), it obeys a modified version of
(\ref{asymp-w-r}), namely (\ref{RS}), where $R_*$ is replaced
with $\widetilde{R}_*$. Strictly speaking, $\widetilde{R}_*$ also
has a mild frequency dependence, but such dependence is not
obvious in the frequency range shown in Fig.~\ref{fig2a}. Second,
$\omega_{\i} R_*$ is no longer a linear function of
$\ln(\omega_{\r} R_*$). The graph $|\omega_{\i} R_*|$ versus
$\ln(\omega_{\r} R_*$) shown in Fig.~\ref{fig2b}  displays a
frequency-dependent slope. The slope of the graph increases from
about 1.3 (1.3) to  about 1.4 (1.9) for the case ${\cal C}=0.306$
(${\cal C}=0.228$) when $j$ goes from 5 to 20. Generally speaking,
the slope demonstrates a greater dependence on frequency for NSs
with smaller compactness. Third, although for realistic stars both
$\omega_\r$ and $\omega_\i$ do not follow (\ref{asymp-w-r}) and
(\ref{asymp-w-i}), the asymptotic behavior Eq.~(\ref{uni}) still
holds. Hence, axial and polar w-modes are located on the same
smooth curve and evenly alternate with each other  in the
high-frequency regime. In the following we try to explain the
above findings in a semi-quantitative way based on the asymptotic
analysis developed for polytropic stars.

\subsection{Physical picture}
First of all, it is instructive to associate such asymptotic
behavior of w-mode QNMs with the physical structure and the EOS of
realistic NSs. As is well known, a typical realistic NS consists
of a core with supernuclear densities (i.e. $\rho \geq 2.8 \times
10^{14}\, {\rm gm}\, {\rm cm}^{-3}$), and a thin crust at
subnuclear densities. The crust can be further divided into an
inner part and an outer part by the neutron drip density $\rho
\approx 4 \times 10^{11}\, {\rm gm}\, {\rm cm}^{-3}$. Due to the
effect of neutron drip, stellar matter softens significantly and
the adiabatic index $\Gamma$, a  typical measure of stiffness
defined by
\begin{equation}\label{adia_index}
\Gamma =
\left(1+\frac{p}{\rho}\right)\frac{\rho}{p}\frac{dp}{d\rho}\,,
\end{equation}
decreases abruptly from about 1.4 to about 0.4 near the neutron
drip density. On the other hand, at the crust-core interface
nuclei disappear and  nuclear matter there consequently becomes
particularly stiff owing to the repulsive part of short range
nuclear force. Correspondingly, $\Gamma$ experiences a sharp rise
from about 1.4 to about 3 around the density $\rho \sim 2 \times
10^{14}\, {\rm gm}\, {\rm cm}^{-3}$. These softening and
stiffening of nuclear matter give rise to huge and rapid changes
in the density profile in the inner crust, which typically has a
thickness of only about a few percent of the stellar radius. For
GWs with wavelengths greater than the thickness of the inner
crust, these drastic changes in the density  effectively lead to
discontinuities in the scattering potential $V$ and its
derivatives as well. Hence, strong reflections of GWs are expected
to occur at the core-crust interface and such a mechanism indeed
attributes to formation of QNMs.

\subsection{Semi-quantitative analysis}
As mentioned above, for a realistic NS, GWs are reflected not only
at the stellar surface  but also from other abrupt changes in
$\rho$ inside the star. In general, the reflection coefficient
$\R$ in (\ref{gfunction}) can be obtained from the leading term in
Eq.~(10) of Ref.~\cite{Berry-1982}, which yields:
\begin{equation}\label{Berry_reflection}
\R  = \frac{\exp(-2i \omega {R}_*)}{4
\omega^2}\int_{0}^{\infty}dr_* \, \frac{dV}{dr_*}\exp(2i \omega
r_*)\,,
\end{equation}
where in $V(r)$ the trivial centrifugal barrier has been excluded.
This equation is generally valid and reduces to (\ref{diff}) if
$V$ (or its $N$-th order derivative) has a step discontinuity at
$r_*=R_*$. It can be clearly seen from (\ref{V-r-star}) that the
potential $V$ is dominated by the density term $\rho$. Hence, the
reflection coefficient $\cal R$ is proportional to the Fourier
transform of ${d\rho}/{dr_*}$.

\begin{figure}[ht]
\centering {
\includegraphics[height=9cm]{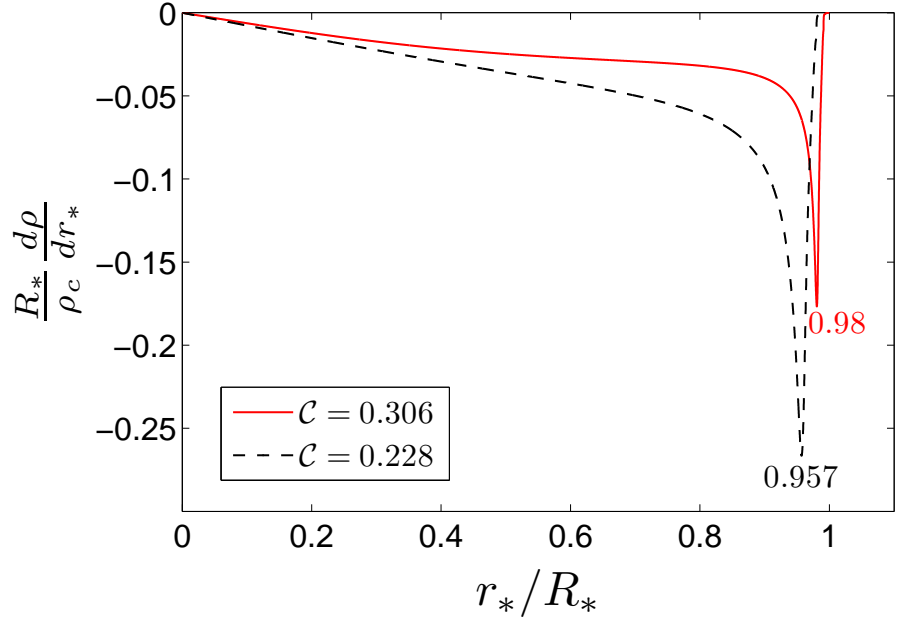}}
\caption{$d \rho /dr_*$ (normalized by $\rho_c/R_*$, with $\rho_c$
being the density at the center of the star) is plotted against
$r_*$ (normalized by $R_*$)
 for two realistic stars obeying the
analytic SLy EOS proposed in Ref.~\cite{anaEOS} with compactness
${\cal C}=0.306,\,0.228$. As labelled in the graphs, $|d \rho
/dr_*|$ attains maximum at $r_*=0.98R_* \,(0.957R_*)$ for the case
${\cal C}=0.306\,(0.228)$. \label{fig3}}
\end{figure}

In Fig.~\ref{fig3} we show ${d\rho}/{dr_*}$  versus $r_*$
 for the two NSs constructed with the
analytic SLy EOS \cite{anaEOS}. It is interesting to see that
there is a sharp peak in $|{d\rho}/{dr_*}|$ at $r_*=R_m$ near the
surface of the star. Generally speaking, $R_m/R_*$ increases
slowly while the width of the peak decreases with increasing
compactness. This peak obviously reflects the abrupt change in the
stiffness of nuclear matter around the core-crust interface. In
fact, it can be argued that $R_m \cong R_2$, where at $r_*=R_2$
the adiabatic index $\Gamma$ equals 2 (see Appendix B and the
data shown in Table~\ref{radii}).

For QNMs with wavelengths much longer than the width of the peak,
${d\rho}/{dr_*}$ can be well approximated by a $\delta$-function.
Following directly from standard theory of asymptotic expansion of
Fourier integrals (see, e.g., \cite{asyexp}), the leading behavior
of the reflection coefficient is
\begin{equation}\label{leading}
\R \sim \frac{a_0(\omega)}{\omega^2}\exp\left[2i \omega
(R_m-{R}_*)\right]\,,
\end{equation}
where $a_0(\omega)$ approaches a constant at low frequencies and
decreases rapidly if $1/|\omega|$ is much smaller than the width
of the peak in $|{d\rho}/{dr_*}|$. Apart from this leading term,
due to the asymmetry in $|{d\rho}/{dr_*}|$ around $r_*=R_m$, there
are other sub-dominant terms in the the reflection coefficient,
leading to the following asymptotic of $\R$:
\begin{equation}
\R \approx \left(\frac{a_{0}}{\omega^{2}}+\frac{a_{1}}{\omega^{3}}
+\frac{a_{2}}{\omega^{4}}+\ldots \right)\exp\left[2i \omega
(R_m-{R}_*)\right]. \label{ref-ana}
\end{equation}
Physically speaking, the coefficients $a_0(\omega)$, $a_1(\omega)$
and $a_2(\omega)$ measure the discontinuities in $\rho$,
${d\rho}/{dr_*}$ and ${d^2\rho}/{dr_*^2}$ around the point
$r_*=R_m$, respectively. As in the case of $a_0(\omega)$,
$a_1(\omega)$ and $a_2(\omega)$ decreases rapidly if $1/|\omega|$
is much smaller than the typical length scale associated with the
corresponding discontinuity.

\subsection{Asymptotic behavior}
Recalling the  asymptotic eigenvalue equation (\ref{eigen}) for
polytropic stars and taking (\ref{ref-ana}) into consideration, we
can show that for  realistic stars the w-mode QNMs are
asymptotically given by the solution of the equation:
\begin{eqnarray}\label{eigen-re1}
\R(\omega) \approx (-1)^{l+1}\exp({-i 2 \omega {R}_m}).
\end{eqnarray}
Since the most singular part of the potential is dominated by the
$\rho$ term, from (\ref{diff}) and (\ref{V-r-star}) we see that
the generalized reflection coefficients for axial and polar
w-modes should have the same frequency dependence but with a sign
difference. Therefore, the two eigenvalue equations for  axial and
polar w-modes can be unified into one single equation:
\begin{eqnarray}
\R_{\rm a}(\omega)\approx (-1)^{l+1+\sigma}\exp({-i 2 \omega_\r
{R}_m})\exp(2 \omega_\i {R}_m) , \label{asymp-w-realistic}
\end{eqnarray}
with  $\R_\a$ being the reflection coefficient of the axial
w-mode, and $\sigma=0\, (-1)$ for the axial (polar) case. In the
asymptotic region, $\R_\a$ is further assumed to have a constant
phase, i.e., $\R_\a(\omega)\approx \exp(i
\theta_\a)|\R_\a(\omega)|$, Eq.~(\ref{asymp-w-realistic}) then
leads to:
\begin{eqnarray}
\omega_\r {R}_m & = & \pi j - \frac{\theta_\a-(l+1+\sigma)\pi}{2}, \label{asymp-w-r-realistic}\\
\omega_\i {R}_m & = & \frac{1}{2}\ln(|\R_\a(\omega)|).
\label{asymp-w-i-realistic}
\end{eqnarray}

Upon equating $\theta_\a-(l+1)\pi$ with $\widetilde{\theta}_b$,
Eq.~(\ref{asymp-w-r-realistic}) agrees nicely with (\ref{RS}),
which is an empirical relation obtained from numerical data such
as those shown in Fig.~\ref{fig2a}, provided that
$\widetilde{R}_*=R_m$. Our argument readily provides the numerical
value of $\widetilde{R}_*$ in (\ref{RS}) a proper physical
interpretation. $\widetilde{R}_*$ is actually  the tortoise
coordinate of the point at which $|d \rho/dr_*|$ attains a
maximum. In addition, $\widetilde{R}_*$ is also close to the
tortoise coordinate of  the point where the adiabatic index is 2
and nuclear matter becomes stiff due to the disappearance of
individual nuclei. This point is confirmed by the numerical data
shown in Table~\ref{radii}.

\begin{figure}[ht]
\centering {
\includegraphics[height=9cm]{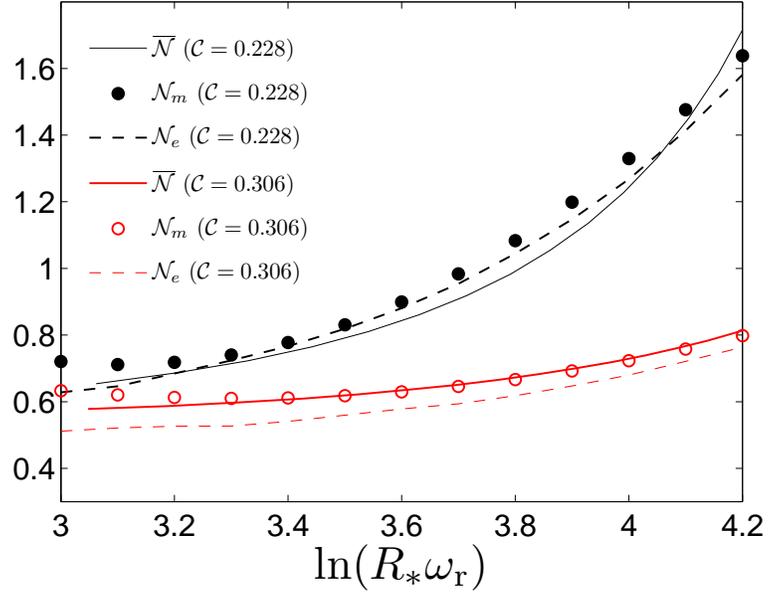}}
\caption{$\overline{{\cal N}}$, ${\cal N}_m$ and ${\cal N}_e$ are
plotted against $\ln ( R_* \omega_\r)$ for two SLy stars with
${\cal C}=0.228, 0.306 $, respectively.\label{fig4}}
\end{figure}

As suggested in (\ref{asymp-w-i-realistic}), $\omega_\i$ depends
crucially on the magnitude of the reflection coefficient
$\R_\a(\omega)$. For the cases of polytropic stars and quark
stars, $\R_\a(\omega) \propto \omega^{-(\N+2)}$, leading directly
to (\ref{asymp-w-i}), and the plot of $|\omega_\i {R}_*|$ versus
$\ln (\omega_\r {R}_*) $ is a straight line with a slope
$m_s=(\N+2)/2 $. In the cases of realistic NS, the polytropic
index $\N$ as well as the adiabatic index $\Gamma$ depend on
density $\rho$. It is therefore reasonable to expect that the plot
of $\omega_\i$ versus $\ln \omega_\r $ is no longer a straight
line. This is in agreement with the result shown in
Fig.~\ref{fig2b}. As a matter of fact, the slope of the curve,
$m_s(\omega_\r)\equiv[\N_m(\omega_\r)+2]/2$, increases gradually
with $\omega_\r$ and the trend is more obvious for the star with
${\cal C}=0.228$. In Fig.~\ref{fig4} the value of $\N_m$ for the
is plotted against $\ln (\omega_\r {R}_*) $ two SLy stars. As
shown in the following discussion, $\N_m$ so obtained could indeed
measure the variation of the polytropic index $\N$ near the
stellar surface.

Since the reflection of GWs mainly takes place around the stellar
surface, we define the average value of $\Gamma$ over a surface
layer with a thickness measured by the wavelength $\lambda =
2\pi/\omega_{\rm r}$, namely:
 \begin{eqnarray}
\overline{\Gamma}=\frac{1}{\Delta}\int_{R_*-\Delta}^{R_*}\Gamma(r_*)dr_*\,,
\end{eqnarray}
where $\Delta= \eta \lambda $ with $\eta=0.61\,(0.7)$ for the
star with ${\cal C}=0.306\,(0.228)$ in our calculations. In
Fig.~\ref{fig4} we compare  the values of $\N_m$ with
$\overline{\N}$ defined by
\begin{eqnarray}
\overline{\N}\equiv\frac{1}{\overline{\Gamma}-1}.
\end{eqnarray}
It is interesting to note that $\overline{\N}$ can indeed
approximate $\N_m$ nicely. Thus, the link between the asymptotic
behavior of w-mode QNMs and the EOS of the nuclear matter
distributed within around one wavelength about the stellar surface
is clearly displayed.

On the other hand, the asymptotic expansion of the reflection
coefficient in (\ref{ref-ana}) could also cast light on the
variation of the slope $m(\omega_\r)$. For GWs with wavelengths
longer than the thickness of the crust, they get reflected from
the narrow peak in $|d \rho/dr_*|$ around the core-crust
interface. Such reflection is measured by the term $a_0/w^2$ in
the expansion of $\cal R$. However, as shown in (\ref{ref-ana}),
there are other contributions $a_1/w^3$ and $a_2/w^4$ due to
reflections resulting from discontinuities of higher orders. Each
of these terms alone leads to an effective slope 1, 1.5 and 2,
respectively. On the other hand,
 for GWs with wavelengths less than the width of the peak,
the value of $a_0$ decreases markedly. Thus, we expect that the
frequency-dependent slope of the graph $|\omega_{\i} R_*|$ versus
$\ln(\omega_{\r} R_*$) in Fig.~\ref{fig2b} increases with
frequency gradually from 1 to 1.5, and then to 2. Such increase in
the slope is expected to be more pronounced if the peak in $d
\rho/d r_*$ is broader. Our conjecture indeed agrees with the
numerical data qualitatively.

Furthermore, we define an effective (frequency-dependent)
polytropic index $\N_e$ by:
\begin{equation}\label{eff_N}
{\cal N}_e=-\frac{\partial \ln |\R_{\rm a}|}{\partial \ln
\omega_{\rm r}}-2\,.
\end{equation}
For polytropic stars $\N_e$ reduces to the standard polytropic
index $\N$. As shown in Fig.~\ref{fig4}, the trend of $\N_e$
(minor oscillations in $\N_e$ have been suppressed) agrees
qualitatively with those of $\N_m$ and $\overline{\N}$. This once
again demonstrates the close relationship between the
high-frequency behavior of w-mode QNMs and the mass distribution
of a NS.

\subsection{Methods relating QNMs}
In spite of the complexity in the situation of realistic NSs, the
relationship between the axial and polar w-modes for these stars
is identical to that for polytropic stars. Due to the sign
difference in the reflection coefficients of axial and polar GWs,
$\omega_\r$ of axial and polar w-mode frequencies, with the same
mode index $j$, differ approximately from each other by $\pi/(2
\widetilde{R}_*)$. Hence, the real part of a polar w-mode
eigenfrequency lies in the middle of two axial w-modes and vice
versa. The difference of $\omega_\r$ leads to the difference in
$\omega_\i$ through (\ref{asymp-w-i-realistic}). Therefore, the
imaginary part of a polar w-mode eigenfrequency should also lies
approximately in the middle of two axial w-modes.

Based on the relationship between the two types of w-mode, two
methods are proposed here to derive approximate polar w-mode
eigenfrequencies from the corresponding axial w-mode
eigenfrequencies. The first method is to approximate $\omega$ of
the $j$-th polar w-mode QNM with the mean value of the
eigenfrequencies of its two neighboring axial w-mode QNMs of the
same $l$, i.e.,
$\omega_j^{(P)}=(\omega_j^{(A)}+\omega_{j+1}^{(A)})/2$. Besides,
from (\ref{asymp-w-r-realistic}) it is obvious that the asymptotic
behaviors of polar w-modes with $l=l_0$ and axial w-modes with
$l=l_0+1$ are  the same. Hence, it is possible to approximate the
frequency of a polar w-mode with $l=l_0$ with that  of an axial
w-mode with the same mode order but $l=l_0+1$. As numerical
calculations of axial w-mode eigenfrequencies are usually much
easier and faster than polar w-mode calculations, the two methods
mentioned above are deemed useful to provide quick ways to locate
approximately polar w-mode QNMs.  As an example to illustrate our
method, we consider the w-mode QNMs of a model C NS (with
compactness $\cc=0.186$ and central density $\rho_c=8.752\times
10^{-4}$) and evaluate approximate polar w-mode frequencies (from
the $5$-th to $14$-th modes) for the case $l_0=2$ using these two
methods. The approximate results so obtained are listed and
compared with the exact values in Table~\ref{Two-Methods}. The
accuracies of the data shown in the table are impressive
especially for high-frequency modes.

\section{Conclusion and discussion}
We study the high-frequency asymptotic behavior of w-model QNMs
of compact stars, including model polytropic stars, quark stars
described by the simplest MIT bag model and NSs constructed with
various realistic EOSs of nuclear matter. By examining relevant
numerical data, we find that there exist some model-independent
generic behaviors in these QNMs. Most interestingly,
notwithstanding the difference between the formalisms of axial
and polar w-modes, the asymptotic behaviors of QNMs of these two
kinds of  w-mode are quite similar. In fact, as shown in
(\ref{uni}), the spectrum of polar w-modes is derivable from its
axial counterpart and vice versa. Such coincidence has motivated
us to study the physical link between these two kinds of w-mode
QNMs. Based on the AAKS formalism for polar oscillations of
compact stars \cite{AAKS98}, we propose another version of ICA
(i.e. $H=0$) and introduce HFA (i.e. $F=0$) to derive a
Klein-Gordon type wave equation (\ref{HFA}) for high-frequency
polar w-modes. Moreover, noting the physical significance of the
density term in the scattering potential, we show that the
potentials in the axial wave equation (\ref{axial-waveeq}) and
the HFA polar wave equation (\ref{HFA}) can be unified and
rewritten into similar forms as in (\ref{V-r-star}) in the
high-frequency regime.

Having established a unified approach to both axial and polar
w-modes, we formulate a WKB analysis to evaluate the
eigenfrequency for relevant QNMs. We note that the crucial factor
affecting the analysis is the discontinuities in the potential (or
the derivatives of the potential). We then apply our theory to the
cases of polytropic stars and quark stars which are described by
smooth EOS. Simple yet accurate formulae (\ref{asymp-w-r}) and
 (\ref{asymp-w-i}), respectively for $\omega_\r$ and $\omega_\i$,  are obtained.
Despite the fact that similar development for realistic NSs is
less trivial and often plagued by the artifacts in EOS, we can
still provide a qualitative explanation for the asymptotic
behavior in such cases. In particular, we point out the
relationships among the asymptotic behavior, the polytropic index
of nuclear matter and the Fourier transform of $d \rho/dr_*$.

It is worthy to note that the asymptotic behavior of QNMs could
easily be
 masked by numerical errors in the density profile. Such
errors could lead to artificial  reflections of GW and hence
unexpected features in the distribution of high-frequency QNMs. In
order to accurately locate high-frequency QNMs, due care has to be
taken in solving the TOV equations for the equilibrium
configuration of a star (see \cite{Zhang_thesis} for a detailed
study for the effect of a tiny truncation error in the density
distribution near the stellar surface on the asymptotic behavior
of QNMs of NSs).

Lastly, we would like to compare the asymptotic behaviors of QNMs
of NSs and BHs. Unlike oscillations of NSs, which can be
classified into fluid and spacetime modes, oscillations of BHs are
all spacetime modes.  Besides, axial and polar mode oscillations
of BH are governed by two different equations, namely the
Regge-Wheeler equation and the Zerilli equation
\cite{RWeq,Zerilli}. However, it is well known that
 the spectra of these two kinds of QNM are identical due to the symmetry in the scattering potentials
 \cite{bhbook,intertwin,qnm_susy}. In fact, the potentials in the
Regge-Wheeler equation and the Zerilli equation are
super-symmetric partners of each other \cite{qnm_susy}. However,
such symmetry between the axial and polar w-modes of NSs no longer
exists.
 Therefore, it is interesting to note that our finding in the
 present paper partially  restores the symmetry between
 these two kinds of QNMs in the sense that their spectra are
 closely related through (\ref{uni}) in the high-frequency regime.
 On the other hand, for QNMs of BHs the frequency of the $j$-th QNM ($j\gg1$) is asymptotically given by (see, e.g.
\cite{maggiore:141301,Berti:2009kk} and references therein):
 \begin{equation}\label{bhasym}
8 \pi M \omega_j \approx \ln 3 -2 \pi i (j+ 1/2)+O(1/{\sqrt{j}})~.
\end{equation}
Therefore, $\omega_\r $ is bounded and approaches a constant,
while $\omega_\i$ is proportional to $j+ 1/2$ and can increase
indefinitely. This is in sharp contrast to (\ref{asymp-w-r}) and
 (\ref{asymp-w-i}), the formulae for NS w-mode QNMs.  Such distinction between QNMs of BHs and NSs is
attributable to the analytic property of relevant potentials in
the wave equation.  For the BH case the potential
 is smooth everywhere for $\infty>r>2M$ and
gravitational waves can escape rapidly, leading to linearly
increasing $|\omega_\i|$. On the other had, as shown in the
present paper, it is the ``singularities" in the density profile
of NSs that give rise to the above-mentioned distinguishing
features in their w-mode spectra.

\begin{acknowledgments}
This work is supported in part by the Hong Kong Research Grants
Council (Grant No: 401807) and the direct grant (Project ID:
2060330) from the Chinese University of Hong Kong. We thank L. M.
Lin  and H. K. Lau for helpful discussions.

\end{acknowledgments}
\appendix
\section{High-order HFA}
From (\ref{delta_p0}) we can expand $F$ in power series of
$1/\omega$, i.e.,
\begin{equation}\label{expansion}
F=F^{(1)}+F^{(2)}+F^{(3)}+\ldots\,
\end{equation} where $F^{(k)}
\sim {\cal O}(S/\omega^k)$ with $k=1,2,3,\ldots$. For purpose of
reference, the lowest four orders of $F^{(k)}$ are listed as
follows:
\begin{eqnarray}
F^{(1)} & = &
-\frac{r e^{-(\nu+\lambda)/2}}{\omega ^2}\frac{dS}{dr_*},\\
F^{(2)} & = & -\frac{m+4\pi r^3 p+(n+1)r}{\omega^2 r}S,\\
F^{(3)} & = & \frac{e^{(\nu-\lambda)/2}}{\omega^4}
              \left\{\frac{4m}{r^2}+8\pi r p+\frac{m+4\pi r^3 p}{r^2}\left[e^{\lambda}(n+1)+
              e^{\lambda}\frac{m-4\pi r^3 \rho}{r}\right]\right\}\frac{dS}{dr_*},\\
F^{(4)}& = &
\frac{e^{\nu}}{\omega^4r^2}\bigg\{\frac{2m}{r}\left(n+1\right)
              +\frac{(m+4\pi r^3 p)^2}{r^2}\left[2+e^{\lambda}(n+1)-\frac{e^{\lambda}}{r}
              \left(4\pi r^3(p+2\rho)-m\right)\right]\nonumber\\
       &   &  +\frac{m+4\pi r^3
       p}{r}\left[\frac{2m}{r}+3(n+1)\right]\bigg\}S.
\end{eqnarray}

Substituting the expansion of $F$ back into (\ref{Allen-St}), we
find a single second-order ODE for polar w-mode oscillation
inside the star:
\begin{eqnarray}
\frac{d^2S}{dr_*^2}+\left(\omega^2-V_{\rm p}\right)S-V_{sf}\left(
F^{(1)}+F^{(2)}+F^{(3)}+\ldots \right)=0\,,\label{S_in}
\end{eqnarray}
where
\begin{eqnarray}
V_{sf} & = & -\frac{4e^{2\nu}}{r^5}\left[\frac{(m+4\pi
pr^3)^2}{r-2m}+4\pi\rho r^3-3m\right]\,.
\end{eqnarray}
We have carried out numerical study to check the validity of
(\ref{S_in}) and found that the second-order approximation (i.e.,
$F=F^{(1)}+F^{(2)}$) can readily reproduce polar w-mode QNMs with
good accuracy and works for both leading and non-leading modes. As
shown in Table~XI,  the accuracy of the perturbation scheme
improves with increasing frequency and increasing order of
expansion as well.
\section{Location of the maximum of
$d\rho/dr_*$} Following directly from the TOV equations
\cite{Tolman:1939jz,Oppenheimer:1939ne}, it can be shown that
\begin{eqnarray}
\frac{d \rho}{d r_*} = -h(r) \frac{(\rho+p)^2}{p},
\end{eqnarray}
where
\begin{eqnarray}
h(r)=\frac{1}{\Gamma(r)}\frac{d r}{d r_*}\frac{m+4 \pi r^3 p}{r
(r- 2m)}
\end{eqnarray}
is a smooth function in comparison  with the term $(\rho+p)^2/p$
near the surface of a NS. Ignoring terms proportional to
$dh/dr_*$, we find that
\begin{eqnarray}
\frac{d^2 \rho}{d r_*^2} \approx -h(r)\frac{\rho+p}{p}\frac{d
\rho}{d r_*}\left[2-\Gamma(r)\right].
\end{eqnarray}
Therefore, the maximum value of $|d\rho/dr_*|$ is attained around
the point $r_*=R_2$ where  $\Gamma=2$.

\newpage
%\bibliography{merge}
\newcommand{\noopsort}[1]{} \newcommand{\printfirst}[2]{#1}
  \newcommand{\singleletter}[1]{#1} \newcommand{\switchargs}[2]{#2#1}

\newpage
\begin{table}[h!]
  \centering
  \caption{The first ten polar w-mode QNMs  of  a polytropic star with EOS $p=100\rho^2$
  and a compactness $0.297$. The first two columns are the exact values of
  $\omega_\r$ and $\omega_\i$. The third and fourth columns are the
  approximate  values of
  $\omega_\r$ and $\omega_\i$ obtained from HFA. All $\omega$ are
  measured in units of $M^{-1}$. The last two columns are
the relative percentage errors in $\omega_\r$ and
$\omega_\i$.}\label{HFA-PS}
\end{table}
\begin{center}
\begin{tabular}{cc|cc|cc}
\hline \hline
 exact $ \omega_\r$    &exact  $|\omega_\i|$  &HFA  $\omega_\r$  & HFA  $|\omega_\i|$  & $|\delta\omega_\r/\omega_\r|$   & $|\delta\omega_\i/\omega_\i|$   \\
\hline
0.5594  &   0.3843 ~ & ~  0.6346  &   0.4642 ~ & ~  13.45  &   20.79  \\
0.4707  &   0.0562 ~ & ~  0.4709  &   0.0282 ~ & ~  0.04   &   49.83  \\
0.6534  &   0.1638 ~ & ~  0.6385  &   0.1248 ~ & ~  2.29   &   23.82  \\
0.8910  &   0.2269 ~ & ~  0.8663  &   0.2084 ~ & ~  2.77   &   8.13   \\
1.1270  &   0.2614 ~ & ~  1.1081  &   0.2527 ~ & ~  1.67   &   3.33   \\
1.3624  &   0.2866 ~ & ~  1.3475  &   0.2814 ~ & ~  1.10   &   1.82   \\
1.5982  &   0.3068 ~ & ~  1.5858  &   0.3033 ~ & ~  0.78   &   1.13   \\
1.8343  &   0.3237 ~ & ~  1.8237  &   0.3212 ~ & ~  0.58   &   0.77   \\
2.0706  &   0.3383 ~ & ~  2.0613  &   0.3364 ~ & ~  0.45   &   0.56   \\
2.3071  &   0.3511 ~ & ~  2.2988  &   0.3497 ~ & ~  0.36   &   0.41   \\
\hline
\end{tabular}
\end{center}
\begin{table}[h!]
  \centering
  \caption{The first ten polar w-mode QNMs  of  a
realistic star with APR2 EOS \cite{APR} and a compactness $0.196$.
The numerical results of $\omega$ are given in units of $M^{-1}$.
The first two columns are the exacts values of
  $\omega_\r$ and $\omega_\i$. The third and fourth columns are the
  approximate  values of
  $\omega_\r$ and $\omega_\i$ obtained from HFA. All $\omega$ are
  measured in units of $M^{-1}$. The last two columns are
the relative percentage errors in $\omega_\r$ and
$\omega_\i$.}\label{HFA-RS}
\end{table}
\begin{center}
\begin{tabular}{cc|cc|cc}
\hline \hline
 exact $ \omega_\r$    &exact  $|\omega_\i|$  &HFA  $\omega_\r$  & HFA  $|\omega_\i|$  & $|\delta\omega_\r/\omega_\r|$   & $|\delta\omega_\i/\omega_\i|$   \\
\hline
0.5050  &   0.2971 ~ & ~  0.4575  &   0.2064 ~ & ~  9.41   &   30.53  \\
0.9209  &   0.3826 ~ & ~  0.8827  &   0.3676 ~ & ~  4.15   &   3.92   \\
1.3067  &   0.4378 ~ & ~  1.2824  &   0.4319 ~ & ~  1.86   &   1.35   \\
1.6840  &   0.4803 ~ & ~  1.6660  &   0.4773 ~ & ~  1.07   &   0.62   \\
2.0565  &   0.5139 ~ & ~  2.0421  &   0.5123 ~ & ~  0.70   &   0.31   \\
2.4266  &   0.5405 ~ & ~  2.4146  &   0.5396 ~ & ~  0.49   &   0.17   \\
2.7966  &   0.5613 ~ & ~  2.7862  &   0.5608 ~ & ~  0.37   &   0.09   \\
3.1681  &   0.5786 ~ & ~  3.1589  &   0.5782 ~ & ~  0.29   &   0.07   \\
3.5413  &   0.5945 ~ & ~  3.5330  &   0.5942 ~ & ~  0.23   &   0.05   \\
3.9155  &   0.6106 ~ & ~  3.9081  &   0.6102 ~ & ~  0.19   &   0.07   \\
\hline
\end{tabular}
\end{center}

\newpage

\begin{table}[h!]
  \centering
  \caption{Relevant physical parameters and values of $b$ of model stars considered in the present paper. Here QS denotes the MIT bag model $p=(\rho-4B)/3$,
  where $B=56 \textrm{MeV fm}^{-3}$ is used throughout the paper. $\cc \equiv M/R$ is the compactness of the star. The $b$ values are calculated with $l=2$.}
  \label{QSNSPB}
\end{table}
\begin{center}
\begin{tabular}{ccccccccc}%{m{2.0cm}m{1.5cm}m{1.5cm}m{1.5cm}m{1.5cm}m{1.5cm}m{1.5cm}}
\hline\hline
          EOS     & $\N$    &   $\cc$  & $R_*$  & $R$    & $M$    & $r_{b}$           & $\theta_{b}$(Axial) & $\theta_{b}$(Polar) \\
\hline QS         & 0       & 0.2644   & 24.617 & 11.292 & 2.986  & $4.3878\times 10^{-4}$  & -$2\pi$             & -$3\pi$             \\
$p=10\rho^2$      & 1       & 0.201    & 5.354  & 2.873  & 0.577  & $3.28 \times 10^{-3}$    & -$2.5\pi$           & -$3.5\pi$           \\
$p=5\rho^{3/2}$   & 2       & 0.156    & 33.602 & 16.961 & 2.646  & $4.07\times10^{-7}$    & -$3\pi$             & -$4\pi$             \\
$p=10\rho^{5/3}$  & 1.5     & 0.195    & 19.277 & 9.147  & 1.784  & $2.24\times10^{-5}$    & -$2.75\pi$          & -$3.75\pi$          \\
$p=10\rho^{11/7}$ & 1.75    & 0.195    & 35.329 & 14.336 & 2.796  & $1.50\times10^{-6}$    & -$2.875\pi$         & -$3.875\pi$         \\
$p=5\rho^{13/9}$  & 2.25    & 0.144    & 53.655 & 23.469 & 3.380  & $2.38\times10^{-8}$    & -$3.125\pi$         & -$4.125\pi$         \\
$p=\rho^{7/5}$    & 2.5     & 0.117    & 13.323 & 6.894  & 0.808  & $2.32\times10^{-6}$    & -$3.25\pi$          & -$4.25\pi$          \\
\hline
\end{tabular}
\end{center}

%\newpage

\begin{table}[h!]
  \centering
  \caption{Theoretical and numerical values of
  $c_1/\pi$, $c_0/\pi$ in (\ref{fitting-quark-r}) ($l=2$) for the model stars tabulated in Table~\ref{QSNSPB}.
    The number enclosed in bracket is the value of $c_0/\pi$
  obtained directly from (\ref{fitting-quark-r}) using the theoretical value of $c_1$ and the
  frequency of the 20-th QNM.   }
 \label{QSNS-C}
\end{table}
\begin{center}
\begin{tabular}{cccccc}
\hline\hline
EOS                &method       & $c_1/\pi$(Axial) & $c_1/\pi$(Polar) & $c_0/\pi$(Axial) & $c_0/\pi$(Polar)  \\
\hline
QS                 &theoretical      & 1                & 1                &  1               & 1.5               \\
                   &numerical        & 1.001            & 1.001            &  0.961(0.981)    & 1.462(1.481)      \\
\hline
$p=10\rho^2$       &theoretical      & 1                & 1                &  1.25            & 1.75              \\
                   &numerical        & 1.002            & 1.002            &  1.164(1.207)    & 1.665(1.708)      \\
\hline
$p=5\rho^{3/2}$    &theoretical      & 1                & 1                &  1.5             & 2                 \\
                   &numerical        & 1.002            & 1.002            &  1.410(1.448)    & 1.911(1.949)      \\
\hline
$p=10\rho^{5/3}$   &theoretical      & 1                & 1                &  1.375           & 1.875             \\
                   &numerical        & 1.002            & 1.002            &  1.293(1.331)    & 1.795(1.831)      \\
\hline
$p=10\rho^{11/7}$  &theoretical      & 1                & 1                &  1.4375          & 1.9375            \\
                   &numerical        & 1.001            & 1.001            &  1.387(1.405)    & 1.887(1.905)      \\
\hline
$p=5\rho^{13/9}$   &theoretical      & 1                & 1                &  1.5625          & 2.0625            \\
                   &numerical        & 1.000            & 1.000            &  1.541(1.535)    & 2.040(2.035)      \\
\hline
$p=\rho^{7/5}$     &theoretical      & 1                & 1                &  1.625           & 2.125             \\
                   &numerical        & 1.001            & 1.001            &  1.563(1.583)    & 2.063(2.083)      \\
\hline
\end{tabular}
\end{center}

\begin{table}[h!]
  \centering
  \caption{Theoretical and numerical values of
  $d_1$, $d_0$ in (\ref{fitting-quark-i}) ($l=2$) for the model stars tabulated in Table~\ref{QSNSPB}.
    The number enclosed in bracket is the value of $d_0$
  obtained directly from (\ref{fitting-quark-i}) using the theoretical value of $d_1$ and the
  frequency of the 20-th QNM.   }
   \label{QSNS-D}
\end{table}
\begin{center}
\begin{tabular}{cccccc}
\hline\hline
EOS                &method       &  $d_1$(Axial) & $d_1$(Polar) & $d_0$(Axial)      & $d_0$(Polar)     \\
\hline
QS                 &theoretical      & -1            & -1           & -0.662            & -0.662           \\
                   &numerical        & -0.997        & -0.997       & -0.677(-0.664)    & -0.678(-0.664)   \\
\hline
$p=10\rho^2$       &theoretical      & -1.5          & -1.5         & -0.343            & -0.343           \\
                   &numerical        & -1.490        & -1.490       & -0.387(-0.347)    & -0.388(-0.347)   \\
\hline
$p=5\rho^{3/2}$    &theoretical      & -2            & -2           & -0.328            & -0.328           \\
                   &numerical        & -2.003        & -2.002       & -0.320(-0.330)    & -0.320(-0.330)   \\
\hline
$p=10\rho^{5/3}$   &theoretical      & -1.75         & -1.75        & -0.175            & -0.175           \\
                   &numerical        & -1.749        & -1.748       & -0.183(-0.177)    & -0.183(-0.177)   \\
\hline
$p=10\rho^{11/7}$  &theoretical      & -1.875        & -1.875       & -0.020            & -0.020           \\
                   &numerical        & -1.890        & -1.892       & 0.045(-0.018)     & 0.052(-0.018)    \\
\hline
$p=5\rho^{13/9}$   &theoretical      & -2.125        & -2.125       & -0.313            & -0.313           \\
                   &numerical        & -2.164        & -2.166       & -0.138(-0.303)    & -0.129(-0.302)   \\
\hline
$p=\rho^{7/5}$     &theoretical      & -2.25         & -2.25        & -0.662            & -0.662           \\
                   &numerical        & -2.281        & -2.279       & -0.525(-0.655)    & -0.533(-0.654)   \\
\hline
\end{tabular}
\end{center}

\newpage

\begin{table}[h!]
  \centering
  \caption{The axial w-mode QNMs ($j=3$ to $j=12$) of a realistic NS (with EOS A
\citep{modelA}, compactness of $0.27$ and central density
$\rho_c=2.227\times 10^{-3}$) are tabulated.  The first two
columns are the exacts values of
  $\omega_\r$ and $\omega_\i$. The third and fourth columns are the
  approximate  values of
  $\omega_\r$ and $\omega_\i$ obtained from the the unified potential in
  (\ref{V-r-star}). All $\omega$ are
  measured in units of $M^{-1}$. The last two columns are
the relative percentage errors in $\omega_\r$ and $\omega_\i$,
  respectively.}\label{Axial-w}
\end{table}
\begin{center}
\begin{tabular}{cc|cc|cc}
\hline\hline
exact $\omega_\r$  & exact $|\omega_\i|$  & $\omega_\r $  &$|\omega_\i| $ & $|\delta\omega_\r/\omega_\r|$   & $|\delta\omega_\i/\omega_\i|$   \\
\hline
1.4031  &   0.3690 ~ & ~  1.4105~~  &   0.3694 ~ & ~  0.52    &   0.13    \\
1.7543  &   0.3974 ~ & ~  1.7604~~  &   0.3973 ~ & ~  0.35    &   0.03    \\
2.1065  &   0.4205 ~ & ~  2.1117~~  &   0.4202 ~ & ~  0.25    &   0.07    \\
2.4593  &   0.4407 ~ & ~  2.4639~~  &   0.4404 ~ & ~  0.18    &   0.07    \\
2.8127  &   0.4593 ~ & ~  2.8167~~  &   0.4589 ~ & ~  0.14    &   0.07    \\
3.1662  &   0.4772 ~ & ~  3.1698~~  &   0.4769 ~ & ~  0.11    &   0.07    \\
3.5195  &   0.4949 ~ & ~  3.5228~~  &   0.4946 ~ & ~  0.09    &   0.06    \\
3.8724  &   0.5128 ~ & ~  3.8754~~  &   0.5125 ~ & ~  0.08    &   0.06    \\
4.2245  &   0.5313 ~ & ~  4.2273~~  &   0.5310 ~ & ~  0.07    &   0.06    \\
4.5752  &   0.5501 ~ & ~  4.5778~~  &   0.5499 ~ & ~  0.06    &   0.05    \\
\hline
\end{tabular}
\end{center}

\begin{table}[h!]
  \centering
  \caption{The polar w-mode QNMs ($j=3$ to $j=12$) of a realistic NS (with EOS A
\citep{modelA}, compactness of $0.27$ and central density
$\rho_c=2.227\times 10^{-3}$) are tabulated.  The first two
columns are the exacts values of
  $\omega_\r$ and $\omega_\i$. The third and fourth columns are the
  approximate  values of
  $\omega_\r$ and $\omega_\i$ obtained from the the unified potential in
  (\ref{V-r-star}). All $\omega$ are
  measured in units of $M^{-1}$. The last two columns are
the relative percentage errors in $\omega_\r$ and $\omega_\i$,
  respectively.}\label{Polar-w}
\end{table}
\begin{center}
\begin{tabular}{cc|cc|cc}
\hline\hline
exact $\omega_\r$    & exact $|\omega_\i|$  & $\omega_\r $  &$|\omega_\i| $ & $|\delta\omega_\r/\omega_\r|$   & $|\delta\omega_\i/\omega_\i|$   \\
\hline
1.2266  &   0.3507 ~ & ~  1.2142~~  &   0.3381 ~ & ~  1.01    &   3.59    \\
1.5782  &   0.3837 ~ & ~  1.5695~~  &   0.3764 ~ & ~  0.55    &   1.90    \\
1.9302  &   0.4093 ~ & ~  1.9235~~  &   0.4044 ~ & ~  0.35    &   1.18    \\
2.2828  &   0.4308 ~ & ~  2.2774~~  &   0.4273 ~ & ~  0.24    &   0.81    \\
2.6359  &   0.4500 ~ & ~  2.6314~~  &   0.4474 ~ & ~  0.17    &   0.59    \\
2.9894  &   0.4682 ~ & ~  2.9856~~  &   0.4661 ~ & ~  0.13    &   0.45    \\
3.3429  &   0.4860 ~ & ~  3.3395~~  &   0.4843 ~ & ~  0.10    &   0.35    \\
3.6960  &   0.5038 ~ & ~  3.6930~~  &   0.5024 ~ & ~  0.08    &   0.29    \\
4.0486  &   0.5220 ~ & ~  4.0459~~  &   0.5208 ~ & ~  0.07    &   0.24    \\
4.4001  &   0.5407 ~ & ~  4.3977~~  &   0.5396 ~ & ~  0.05    &   0.20    \\
\hline
\end{tabular}
\end{center}

\newpage

\begin{table}[h!]
  \centering
  \caption{The values of $R_m/R_*$, $R_2/R_*$ and $\widetilde{R}_*/R_*$ for the two realistic stars constructed
  with the analytic  approximation of the
SLy EOS \cite{anaEOS} with compactness ${\cal C}=0.306,\,0.228$
are tabulated. The value of $\widetilde{R}_*$ is obtained from
from the best linear fit to the data shown in Fig.~\ref{fig2a}.
}\label{radii}
\end{table}
\begin{center}
\begin{tabular}{c|ccccc}
\hline \hline ${\cal C}$ & $R_m/R_*$   & $R_2/R_*$  & $\widetilde{R}_*/R_*$ \\
\hline
0.306 & 0.980 & 0.978 & 0.984 \\
0.228 & 0.957 & 0.950 & 0.958 \\ \hline
\end{tabular}
\end{center}

\begin{table}[h!]
  \centering
  \caption{Evaluation of $l=2$ polar w-modes from the corresponding axial
  w-modes for  a model C NS (with compactness $\cc=0.186$ and central density
$\rho_c=8.752\times 10^{-4}$). The
  first two columns show the exact values of $\omega_\r$ and
  $\omega_\i$. The third and the fourth columns are respectively
  approximate values of $\omega_\r$ and
  $\omega_i$ obtained from (\ref{uni}). The last two columns are
  the approximate values of $\omega_\r$ and
  $\omega_\i$ obtained from corresponding axial w-modes with $l=3$. All $\omega$ are
  measured in units of $M^{-1}$.}\label{Two-Methods}
\end{table}
\begin{center}
\begin{tabular}{cc|cc|cc}
\hline\hline
exact $\omega_r$    & exact $|\omega_i|$  & $\omega_r$  &  $|\omega_i|$ &  $\omega_r$  &  $|\omega_i|$   \\
\hline
0.9135  &   0.2286  &   0.9134~~  &   0.2283~~  &   0.9118~~  &   0.2287  \\
1.0796  &   0.2398  &   1.0795~~  &   0.2396~~  &   1.0781~~  &   0.2399  \\
1.2453  &   0.2493  &   1.2453~~  &   0.2491~~  &   1.2440~~  &   0.2494  \\
1.4110  &   0.2574  &   1.4110~~  &   0.2573~~  &   1.4099~~  &   0.2575  \\
1.5769  &   0.2645  &   1.5769~~  &   0.2644~~  &   1.5758~~  &   0.2646  \\
1.7429  &   0.2711  &   1.7430~~  &   0.2710~~  &   1.7420~~  &   0.2711  \\
1.9091  &   0.2773  &   1.9091~~  &   0.2772~~  &   1.9083~~  &   0.2773  \\
2.0754  &   0.2831  &   2.0754~~  &   0.2831~~  &   2.0746~~  &   0.2832  \\
2.2419  &   0.2888  &   2.2419~~  &   0.2887~~  &   2.2411~~  &   0.2888  \\
2.4085  &   0.2943  &   2.4086~~  &   0.2943~~  &   2.4078~~  &   0.2943  \\
\hline
\end{tabular}
\end{center}
\newpage

\end{document}